\documentclass[twocolumn,aps,prd,preprintnumbers,nofootinbib]{revtex4}
\usepackage[dvips]{graphicx}
\usepackage{bm,latexsym,amsmath,amssymb,amsfonts}
\usepackage[normalem]{ulem}

\begin{document}

\title{\uline{}Quasi-local characteristics of dynamical extreme black holes}

\author{Tetsuya Shiromizu${}^{1}$, Sumio Yamada${}^{2}$ and Kentaro Tanabe${}^{3}$}
\affiliation{${}^{1}$Department of Physics, Kyoto University, Kyoto 606-8502, Japan}
\affiliation{${}^{2}$Department of Mathematics, Gakushuin University, Tokyo 171-8588, Japan}
\affiliation{${}^{3}$Departament de F{\'\i}sica Fonamental, Institut de Ci\`encies del Cosmos, 
Universitat de Barcelona, Mart\'{\i} i Franqu\`es 1, E-08028 Barcelona, Spain}

\begin{abstract}
Introducing the concept of the extreme trapping horizon, we discuss geometric features of 
dynamical extreme black holes in four dimensions and then 
derive the integral identities which hold for the dynamical extreme black holes. 
We address the causal/geometrical features too. 
\end{abstract}
\maketitle

%\section{Introduction}

%\section{Momentarily static case}

\section{Introduction} 

Recently it has been shown that the extreme Reissner-Nordstrom black holes are unstable under linear 
perturbations \cite{Aretakis:2011ha, Lucietti:2012xr} 
(See also Refs. \cite{Aretakis2, Lucietti:2012sf, Murata:2012ct} for the extreme Kerr spacetime).  
 To show this, the conservation feature 
under perturbations played a central role, called Aretakis' constant. Using coordinate 
transformations, one may
identify the horizon with a set in an artificial null infinity and then the conservation law can be written 
as an asymptotic  quantity conserved on the artificial null infinity \cite{Bizon:2012we}. But, the 
meaning of  those quantities remains  unclear. 
In Ref.~\cite{Murata:2013daa}, using non-linear numerical analysis of the back-reaction to 
the spacetimes with the spherical symmetry,  dynamical extreme black holes 
under a specific initial condition has been created.

In this paper, we will discuss the features of the dynamical extreme black holes and we will define extreme trapping horizon in doing so. 
We will present a pair of integral formulas which are valid on 
2-surfaces corresponding to horizons of the extreme black holes. We will derive them in two 
different forms. The tool we use is the second 
variation of the area of minimal surface, as in 
\cite{Hawking:1971vc,Galloway:2005mf}(See also Refs.~\cite{Dain:2011kb,Dain:2013qia}
where the same observation was used to discuss an inequality for charged black holes.)  
The integral identities presented in this 
article may give us extra information for understanding the origin of  Aretakis' constant, as both observations rely on the extremality of the 
black hole spacetimes. For the moment, however,  we cannot  
find direct relations between them. We also discuss some general features of the extreme trapping horizon. 

The article is organized as follows. In Sec.II, we derive a surface integral 
identity on momentarily static slices. Such surfaces correspond to the event horizon of the extreme Reissner-Nordstrom 
black hole, for example. In Sec.III, we give a definition of the extreme trapping horizon in a general set-up 
and another surface integral 
identity. We also address the properties of the extreme trapping horizon. Finally we will give the summary 
and discussion in Sec.IV. 

\section{dynamical extreme black holes in momentarily static slices}

Let us consider a 3-dimensional spacelike hypersurface $\Sigma$ and a compact
2-dimensional submanifold $S$ in $\Sigma$. 
The following Riemannian geometric identity then holds unconditionally,   
%========<Equation>========%
%
\begin{eqnarray}
\mbox \pounds _r k=-\varphi^{-1}{\cal D}^2 \varphi +\frac{1}{2}{}^{(2)}R-\frac{1}{2}
\Bigl({}^{(3)}R+k^2+k_{ab}k^{ab} \Bigr),\label{rgi}
\end{eqnarray}
%
%==========================%
where $r^a$ is the unit outward normal vector of $S$ in $\Sigma$, $\varphi$ is the lapse function, 
$k_{ab}$ is the extrinsic curvature of the surface $S$ in $\Sigma$ and ${}^{(2)}R$ is the Ricci scalar 
of $S$. ${\cal D}_a$ is the covariant derivative with respect to the induced metric of $S$. Recall that this identity is closely related to the second variation formula of the area functional of $S \subset \Sigma$.

In general the expansion rate of the null geodesic congruence is given by 
%========<Equation>========%
%
\begin{eqnarray}
\theta = k+K_{ab}r^a r^b+K,
\end{eqnarray}
%
%==========================%
where $K_{ab}$ is the extrinsic curvature of $\Sigma$ and $K$ is the trace part of $K_{ab}$. If one 
considers the momentarily static (or, equivalently, time symmetric) slice $\Sigma$, that is, the extrinsic 
curvature of $\Sigma$ vanishes ($K_{ab}=0$), 
the vanishing of $\theta$ corresponds to the vanishing of $k$, so that $S$ is a minimal surface in $\Sigma$. 
The surface where the 
expansion vanishes is defined as the cross section of the 
trapping horizon, which is nearly the same as the apparent horizon 
\cite{Hayward:1993tt}. 
The first variation of $k$ does not vanish in general. In the dynamical extreme 
black holes, however, 
we consider the ``horizon" $S_H \subset \Sigma$ where the first variation of the 
expansion also vanishes as
%========<Equation>========%
%
\begin{eqnarray}
\mbox \pounds_r k=0.
\end{eqnarray}
%
%==========================%
We may call this surface $S_H$ the extreme minimal surface. 
This definition of the ``horizon" is consistent  with the one
given by Israel \cite{Israel}. Note that the trapped surfaces do not exist inside of the 
horizon $S_H$. 
Then the surface integral on $S_H$ of Eq. (\ref{rgi}) implies 
%========<Equation>========%
%
\begin{eqnarray}
\int_{S_H} {}^{(2)}RdS = \int_{S_H}\Bigl({}^{(3)}R+k_{ab}k^{ab}+2\varphi^{-2}({\cal D}\varphi)^2 \Bigr)
dS. 
\end{eqnarray}
%
%==========================%
On the momentarily static slices, the Hamiltonian constraint becomes 
%========<Equation>========%
%
\begin{eqnarray}
{}^{(3)}R=2T_{ab}n^an^b,
\end{eqnarray}
%
%==========================%
where $T_{ab}$ is the energy-momentum tensor and $n^a$ is the future directed unit normal vector of 
the current time slice $\Sigma$. If the energy condition is satisfied so that $T_{ab}n^an^b >0$ (the 
inequality is strict because we consider the extreme Reissner-Nordstrom-type black hole and there is 
always a non-trivial 
contribution from the Maxwell 
field to the energy-momentum tensor), 
the right-hand side is positive. Then, the Gauss-Bonnet theorem tells us the left-hand side becomes 
$8\pi$. We then have 
%========<Equation>========%
%
\begin{eqnarray}
\int_{S_H}\Bigl(2T_{ab}n^a n^b+k_{ab}k^{ab}+2\varphi^{-2}({\cal D}\varphi)^2 \Bigr)dS=8\pi. 
\end{eqnarray}
%
%==========================%
This is an integral identity which holds on any momentarily static slices. Note that we cannot 
apply it to the static slice of the extreme Reissner-Nordstrom black hole because the static slice 
crosses the bifurcation surface and $n^a$ cannot be kept to be a timelike vector field. 

One may, however,  apply the above equality to cases with 
the massless scalar fields on the dynamical extreme black hole. 
In particular, the identity in the spherically symmetric case gives the following identity;
%========<Equation>========%
%
\begin{eqnarray}
\int_{S_H}g^{rr} \Bigl(2E_r^2+{\phi'}^2 \Bigr)dS = 8 \pi, \label{msid}
\end{eqnarray}
%
%==========================% 
where $E_r=F_{ra}n^a$ is the radial component of the electric field, $\phi$ is a massless scalar field 
and the prime stands for the derivative along the radial direction and $g^{rr}$ is determined 
by solving the Hamiltonian constraint. Here 
we have used that momentary staticity implies $\dot \phi=0$ on $\Sigma$ 
(the dot stands for the time derivative) through the momentum constraint. 

When there is a sequence of the momentarily static initial data, which suggests an underlying time 
evolution, the left-hand side in Eq. (\ref{msid}) 
can be regarded as a conserved quantity. We remark that this is consistent with the fact that 
$\phi'$ becomes constant at a later time in the linear 
perturbation level \cite{Aretakis:2011ha,Lucietti:2012xr}.

\section{General cases} 

In this section, we will consider more general cases. We will define the extreme trapping horizon and 
give the surface integral identity at the extreme trapping horizon. We also discuss the causal/geometrical 
properties. 

\subsection{Extreme trapping horizon}

We first derive the basic equation,  and then define the extreme trapping horizon.   
Here we will employ the double null coordinate. 
In the double null decomposition, the metric of spacetimes is written as 
$g_{ab}=h_{ab}-e^{-f} (n_{+a}n_{-b}+n_{-a}n_{+b})$, where $n_\pm$ are outgoing/ingoing null vectors. 
The 2-surfaces with the induced metric $h_{ab}$ 
have the canonical parameters $(\xi_+,\xi_-)$, so that $n_\pm^a=
e^f((\partial_{\xi_\pm})^a-r^a_\pm)$ with $r^a_\pm$ is the shift vector. Then 
we have \cite{Hayward:1993tt}
%========<Equation>========%
%
\begin{eqnarray}
& & e^f \mbox \pounds_- \theta_++e^f \theta_+ \theta_-+\frac{1}{2}{}^{(2)}R-\tau_a \tau^a
-{\cal D}_a \tau^a \nonumber \\
& & ~~=e^{-f} T_{ab}n^a_+ n^b_-, \label{cf}
\end{eqnarray}
%
%==========================%
where $\theta_\pm=(1/2) h^{ab}\mbox \pounds_\pm h_{ab}$, 
$\mbox \pounds_\pm $ is the Lie derivative with respect to 
$e^{-f}n^a_\pm$ and $\tau_a=\omega_a-{\cal D}_af/2$. ${\cal  D}_a$ is the 
covariant derivative with respect to $h_{ab}$ and $\omega_a=
(1/2)e^{-f}h_{ab}(n_+^c \nabla_c n_-^b-n_-^c \nabla_c n_+^b )$. 

If 
%========<Equation>========%
%
\begin{eqnarray}
\theta_+=e^f \mbox \pounds_- \theta_+=0 \label{eth}
\end{eqnarray}
%
%==========================%
is satisfied on a $S_{\rm eth}$, we call $S_{\rm eth}$ the 
{\it extreme trapping horizon}\footnote{Originally the definition of the trapping horizon has the 
form of the ``time development" \cite{Hayward:1993tt}.}. In addition, inspired by the model of 
extreme Reissner-Nordstrom spacetime and the recent numerical study \cite{Murata:2013daa}, we require 
that the region inside $S_{\rm eth}$ are not {\em trapped} in the definition. 

We are interested in 
the geometrical feature of the time ``development" ${\cal H}_{\rm ex}:=\cup_{t \in R}S_{\rm eth}(t)$ 
of $S_{\rm eth}$. Let us suppose $z$ to 
be the tangent vector of ${\cal H}_{\rm ex}$ written as $z=e^{-f}(\alpha n_+ +\beta n_-)$. 
From the definitions, 
$\mbox \pounds_z \theta_+|_{{\cal H}_{\rm ex}}=
(\alpha \mbox \pounds_{+}\theta_+ +\beta \mbox \pounds_{-}\theta_+)|_{{\cal H}_{\rm ex}} =0$. 
The last equality comes from the definition of ${\cal H}_{\rm ex}$. 
Since $\mbox \pounds_{-}\theta_+|_{{\cal H}_{\rm ex}} =0$, we see that $\alpha=0$ or 
$\mbox \pounds_{+}\theta_+|_{{\cal H}_{\rm ex}}=0$ holds. 
In the former case, $z=\beta e^{-f}n_-$. This cannot be the case because $z \propto e^{-f}n_+$ for the 
future extreme trapping horizon, which is the future event horizon, of a certain domain of outer 
communications in the extreme Reissner-Nordstrom spacetime. 
Then, it is natural to expect the latter, namely $\mbox \pounds_{+}\theta_+|_{{\cal H}_{\rm ex}}=0$, 
holds for the 
extreme trapping horizon. This would imply a strong constraint on the induced geometry of 
${\cal H}_{\rm ex}$ and matters, that is, the shear $\sigma_{ab}^+$ of 
outgoing null geodesic congruences and $T_{ab}n^a_+n^b_+$ vanish on ${\cal H}_{\rm ex}$. Indeed, while we do not have 
 the causal feature of ${\cal H}_{\rm ex}$ unlike the case for the non-extremal cases, we could instead show
the presence of the Killing vector along ${\cal H}_{\rm ex}$, from which the symmetry of the extreme 
trapping horizon would follow; a strong geometric consequence.

\subsection{The surface integral at the extreme trapping horizon}

Integrating Eq. (\ref{cf}) over the extreme trapping horizon gives us 
%========<Equation>========%
%
\begin{eqnarray}
4\pi = \int_{S_{\rm eth}} \Bigl( e^{-f} T_{ab}n^a_+ n^b_-+\tau_a \tau^a \Bigr)dS. \label{nullcon}
\end{eqnarray}
%
%==========================%
When the surface develops such that Eq. (\ref{eth}) holds, the above quantity is conserved.

As before, we consider cases with the massless scalar fields on spherical symmetric and 
dynamical extreme black hole. The energy-momentum tensor is $T_{ab}^{({\rm scalar})}=
\partial_a \phi \partial_b \phi -(1/2)g_{ab}(\partial \phi )^2$. 
In the double null coordinate, the metric is 
%========<Equation>========%
%
\begin{eqnarray}
ds^2=-2e^{-f(u,v)}dudv+(r(u,v))^2d\Omega^2.
\end{eqnarray}
%
%==========================%
Then we see $T^{({\rm scalar})}_{ab}n^a_+n^b_- \propto T^{({\rm scalar})}_{uv}=0$. 
Therefore, there is no contribution of the probe massless scalar fields 
into Eq. (\ref{nullcon}). Hence, in this setting, it is unlikely that Eq. (\ref{nullcon}) reflects 
Aretakis' constant. 

On the other hand, the Maxwell field has the contribution to the integral (\ref{nullcon}) for the 
extreme Reissner-Nordstrom solution, and we can check that 
%========<Equation>========%
%
\begin{eqnarray}
4\pi = \int_{S_{\rm eth}} e^{-f} T_{ab}^{({\rm Maxwell})}n^a_+n^b_-dS
\end{eqnarray}
%
%==========================%
holds, where $T_{ab}^{({\rm Maxwell})}=2(F_{ac}F_{b}^{~c}-\frac{1}{4}g_{ab}F^2)$. 
We used the fact that $\tau^a$ vanishes for spherical symmetric spacetimes. 

\subsection{Properties of extreme trapping horizon}

There is an inequality $A_{\rm eth} \geq 4\pi Q^2$, where  $A_{\rm eth}$ 
is the area of extreme trapping horizon and $Q$ is the charge defined on $S_{\rm eth}$
($Q=\frac{1}{4\pi}\int_{S_{\rm eth}}F$), as shown in Ref. \cite{Dain:2011kb}. On the 
other hand, it is reported in the numerical study that the area of the event horizon ($A_{\rm EH}$) 
is less than $4\pi Q^2$, $A_{\rm EH} \leq 4\pi Q^2$ \cite{Murata:2013daa}. 
Note that the numerical study is restricted to the spherically symmetric case. 
In order to understand these mutually incompatible inequalities (usually we expect 
$A_{\rm eth} \leq A_{\rm EH}$), we shall consider three possible cases separately: (i) 
the extreme trapping horizon is inside of the event horizon or (ii) the extreme trapping horizon is 
outside of the event horizon or (iii)the extreme trapping horizon coincides with the event 
horizon. Of course, the case (ii) is unlikely, but it is non-trivial to 
remove it if the extreme trapping horizon indeed exists. 

In the case (i), if the spacetime becomes to be stationary at a sufficiently later time, 
the extreme trapping horizon will approach the event horizon. Then we see that the 
expansion rate of the null geodesic congruences should be negative there because of the inequalities 
$A_{\rm EH} \leq 4\pi Q^2 \leq A_{\rm eth}$. 
Since there are not trapped surfaces in the current cases, this is impossible. 
Therefore, the extreme trapping horizon will not approach the event 
horizon if $A_{\rm EH} \leq 4\pi Q^2$ holds. This observation is consistent with the existing 
results in the numerical study \cite{Murata:2013daa}. 

The case (ii) is possible in principle if the extreme trapping horizon exists. However, one thinks 
that the perturbations would easily destroy the extreme trapping horizon by the following 
argument \footnote{The similar argument to show that trapped surfaces are inside of the 
event horizon if the trapped surfaces exist.}.  
Indeed, it is easy to realize situations with non-zero shear $\sigma_{ab}^+$ (we suppose that 
the generic condition \cite{HE} is satisfied) near the extreme trapping horizon so that  there 
is a 2-surface with a very small yet  positive expansion. Then the Raychauduhri 
equation tells us that the expansion will be negative at a sufficiently later time due to the presence 
of the shear. 
This means the formation 
of the trapped surface outside of the event horizon. Following the standard argument on black holes \cite{Hawking:1971vc}, 
this is impossible if the cosmic censorship 
conjecture holds \cite{Penrose:1969pc}.

The case (iii) is expected to be the stationary or static in asymptotically flat spacetimes. 
Then the spacetime will be the extreme Reissner-Nordstrom one if the black hole is single and non-rotating 
one. 

There are two remarks. 
In a reality, the case (i) is more probable than (ii). This is because the perturbations 
which induce a nontrivial shear seem to exist in general except for spherical symmetric cases 
\footnote{Note that the shear 
vanishes if one considers the spherical symmetric cases, and then the case (ii) may occur.}. 
In the case (i), it is expected that the trapped surface is easily 
formed by the same argument as in the case (ii) and it then contradicts the presence of the extreme trapping horizon. 
This remains a speculation, however, as we cannot be certain of the presence of trapped surface inside of the event horizon due to the possible spacetime singularities, which could precede the formation of 
trapped surfaces.

\section{Summary and discussion}

We derived a pair of surface integral identities on the extreme minimal 
surface and the extreme trapping horizon in the dynamical extreme black hole geometry. They may give us a new 
constraint on the dynamical extreme black holes. 
For example, we can see that the area of the extreme trapping horizon is larger than $4\pi Q^2$, where 
$Q$ is the total charge defined at the extreme trapping horizon. 
We also examined the general feature of the 
extreme trapping horizon. Then it was shown that the shear of null geodesic congruence and a component 
of the energy-momentum tensor vanish on the extreme trapping horizon. From these facts it follows that
the extreme trapping horizon has the symmetry. 
And we observed that the extreme trapping horizon seems to be 
inside of the event horizon and both horizons will not approach each other even at a 
sufficiently later time if the area of the cross section of the event horizon is less than $4\pi Q^2$
(this is indicated through the recent numerical study \cite{Murata:2013daa}). 
This means that the spacetime could be dynamical forever. We also had the comment on the 
linear instability founded 
recently. However, we cannot show the direct relation between our integral identities and the 
conserved quantities of the perturbations on the extreme black holes.  Since the 
extremality of the geometry is common in both settings, one may hope that they are related 
somehow. Nevertheless, our identities will be useful for the estimation of the error of 
numerical study. 

Finally we list a set of remaining issues and open questions.  Firstly, the geometrical/causal aspects of the 
extreme trapping horizon should be  further clarified. 
Next, our argument relied on the two-dimensionality as we used the Gauss-Bonnet theorem, while the instability of the extreme black hole has been observed
regardless of the spacetime dimensions \cite{Murata:2012ct}. 
Thus  we pose the problem of extending our argument to higher dimensions. Finally, it is natural to consider the extreme trapping horizon corresponding to the extreme Kerr black hole (See Ref. 
\cite{Jaramillo:2012zi}), where instead of charge, the contribution 
of the angular momentum needs to be taken into account.

\begin{acknowledgments}
We thank RIMS, Kyoto University, where 
this work was initiated during the workshop on ``Explorations in Differential Geometry of Submanifold". 
TS and SY are supported by Grant-Aid for Scientific Research from Ministry of Education, Science,
Sports and Culture of Japan (Nos.~21244033 and 25610055(TS), Nos.~23654061 and 24340009(SY)). 
KT was supported by a grant for research abroad by JSPS. 
\end{acknowledgments}

%---------   References   ---------%

%---------   References   ---------%


\begin{thebibliography}{99}



%\cite{Aretakis:2011ha}
\bibitem{Aretakis:2011ha} 
  S.~Aretakis,
  %``Stability and Instability of Extreme Reissner-Nordstr\'om Black Hole Spacetimes for Linear Scalar Perturbations I,''
  Commun.\ Math.\ Phys.\  {\bf 307}, 17 (2011);
  %%CITATION = ARXIV:1110.2007;%%
Annales Henri Poincare {\bf 12}, 1491 (2011).
  %%CITATION = ARXIV:1110.2009;%%

\bibitem{Lucietti:2012xr} 
  J.~Lucietti, K.~Murata, H.~S.~Reall and N.~Tanahashi,
  %``On the horizon instability of an extreme Reissner-Nordstr\'om black hole,''
  JHEP {\bf 1303}, 035 (2013)
  [arXiv:1212.2557 [gr-qc]].
  %%CITATION = ARXIV:1212.2557;%%



\bibitem{Aretakis2}
S.~Aretakis,~
J.\ Funct.\ Anal.\  {\bf 263}, 2770 (2012);
  %%CITATION = ARXIV:1110.2006;%%
 arXiv:1206.6598 [gr-qc].
  %%CITATION = ARXIV:1206.6598;%%

\bibitem{Lucietti:2012sf} 
  J.~Lucietti and H.~S.~Reall,
  %``Gravitational instability of an extreme Kerr black hole,''
  Phys.\ Rev.\ D {\bf 86}, 104030 (2012)
  [arXiv:1208.1437 [gr-qc]].
  %%CITATION = ARXIV:1208.1437;%%

\bibitem{Murata:2012ct} 
  K.~Murata,
  %``Instability of higher dimensional extreme black holes,''
  Class.\ Quant.\ Grav.\  {\bf 30}, 075002 (2013)
  [arXiv:1211.6903 [gr-qc]].

%\cite{Bizon:2012we}
\bibitem{Bizon:2012we} 
  P.~Bizon and H.~Friedrich,
  %``A remark about wave equations on the extreme Reissner-Nordstr\'om black hole exterior,''
  Class.\ Quant.\ Grav.\  {\bf 30}, 065001 (2013)
  [arXiv:1212.0729 [gr-qc]].
  %%CITATION = ARXIV:1212.0729;%%

\bibitem{Murata:2013daa} 
  K.~Murata, H.~S.~Reall and N.~Tanahashi,
  %``What happens at the horizon(s) of an extreme black hole?,''
  arXiv:1307.6800 [gr-qc].
  %%CITATION = ARXIV:1307.6800;%%

%\cite{Hawking:1971vc}
\bibitem{Hawking:1971vc} 
  S.~W.~Hawking,
  %``Black holes in general relativity,''
  Commun.\ Math.\ Phys.\  {\bf 25}, 152 (1972).
  %%CITATION = CMPHA,25,152;%%
  %470 citations counted in INSPIRE as of 06 Jul 2013

%\cite{Galloway:2005mf}
\bibitem{Galloway:2005mf} 
  G.~J.~Galloway and R.~Schoen,
  %``A Generalization of Hawking's black hole topology theorem to higher dimensions,''
  Commun.\ Math.\ Phys.\  {\bf 266}, 571 (2006)
  [gr-qc/0509107].
  %%CITATION = GR-QC/0509107;%%


\bibitem{Dain:2011kb} 
  S.~Dain, J.~L.~Jaramillo and M.~Reiris,
  %``Area-charge inequality for black holes,''
  Class.\ Quant.\ Grav.\  {\bf 29}, 035013 (2012).

%\cite{Dain:2013qia}
\bibitem{Dain:2013qia} 
  S.~Dain, M.~Khuri, G.~Weinstein and S.~Yamada,
  %``Lower Bounds for the Area of Black Holes in Terms of Mass, Charge, and Angular Momentum,''
Phys.\ Rev.\ D {\bf 88}, 024048 (2013) 
[arXiv:1306.4739 [gr-qc]].

\bibitem{Hayward:1993tt} 
  S.~A.~Hayward,
  %``General laws of black hole dynamics,''
  Phys.\ Rev.\ D {\bf 49}, 6467 (1994);
  %%CITATION = PHRVA,D49,6467;%%
 S.~A.~Hayward, T.~Shiromizu and K.~-i.~Nakao,
  %``A Cosmological constant limits the size of black holes,''
  Phys.\ Rev.\ D {\bf 49}, 5080 (1994)
  [gr-qc/9309004].
  %%CITATION = GR-QC/9309004;%%

\bibitem{Israel}
W. Israel, Phys. Rev. Lett., {\bf 57},397 (1986). 

\bibitem{HE}
S. W. Hawking and G. F. R. Ellis, {\it ``Large scale structure of space-time"}, 
Cambridge University Press(Cambridge, , 1973).

\bibitem{Penrose:1969pc} 
  R.~Penrose,
  %``Gravitational collapse: The role of general relativity,''
  Riv.\ Nuovo Cim.\  {\bf 1}, 252 (1969)
  [Gen.\ Rel.\ Grav.\  {\bf 34}, 1141 (2002)].
  %%CITATION = RNCIB,1,252;%%

\bibitem{Jaramillo:2012zi} 
  J.~L.~Jaramillo,
  %``A note on degeneracy, marginal stability and extremality of black hole horizons,''
  Class.\ Quant.\ Grav.\  {\bf 29}, 177001 (2012)
  [arXiv:1206.1271 [gr-qc]].
  %%CITATION = ARXIV:1206.1271;%%

\end{thebibliography}
\end{document}